\documentclass[usenatbib,usegraphicx]{mn2e}
\newcommand{\Ledd}{$L/L_{\rm Edd}$}
\newcommand{\Mbh}{$M_{\rm BH}$}
\newcommand{\Hb}{H$\rm\beta$}
\newcommand{\Ox}{[\mbox{O\,{\sc iii}}]}
\newcommand{\C}{\mbox{C\,{\sc iv}}}
\newcommand{\Fe}{\mbox{Fe\,{\sc ii}}}
\newcommand{\fnrepeat}[1]{$^{\ref{#1}}$}

\title[On the origin of the \C\ Baldwin effect in AGN]{On the origin of the \C\ Baldwin effect in AGN}
\author[Baskin \& Laor]{Alexei Baskin\thanks{E-mail:alexei@physics.technion.ac.il(AB); laor@physics.technion.ac.il (AL)} \& Ari Laor\footnotemark[1] \\
Physics Department, Technion, Haifa~32000,
Israel}
\begin{document}
\maketitle
\begin{abstract}
The origin of the luminosity dependence of the equivalent width
(EW) of broad emission lines in AGN (the Baldwin effect) is not
firmly established yet. We explore this question for the broad
\C~$\lambda 1549$ line using the Boroson \& Green sample of the 87
$z\le 0.5$ Bright Quasar Survey (BQS) quasars. Useful UV spectra
of the \C\ region are available for 81 of the objects, which are
used to explore the dependence of the \C~EW on various emission
properties. We confirm earlier results on the strong correlations
of the \C~EW with some of the emission parameters which define the
Boroson \& Green Eigenvector 1, and with the optical to X-ray
slope $\alpha_{ox}$. In addition, we find a strong correlation of
the \C~EW with the relative accretion rate, \Ledd. Since \Ledd\
drives some of the Eigenvector 1 correlations, it may be the
primary physical parameter which drives the Baldwin effect for \C.
\end{abstract}

\section{Introduction}
The inverse correlation of the broad emission line equivalent
width (EW) with luminosity in AGN, discovered by Baldwin (1977)
for the \C~$\lambda 1549$ line, was intensively explored over the
past 20 years (see a comprehensive review by Osmer \& Shields
1999; and more recent studies by Wilkes et al. 1999; Green et al.
2001; Croom et al. 2002; Dietrich et al. 2002; Kuraszkiewicz et
al. 2002; and Shang et al. 2003). The physical origin for this
effect is not clearly established yet, but a plausible explanation
is softening of the ionizing continuum shape, and increasing gas
metalicity with increasing luminosity (Korista, Baldwin \& Ferland
1998). The softening of the ionizing continuum shape with
luminosity is supported by: 1. some observed
correlations of the emission line strength with the ionizing
spectral shape (Green 1998; Wang, Lu \& Zhou 1998), 2. predictions
of simple accretion disk models for the ionizing spectral shape
dependence on luminosity (Netzer, Laor \& Gondhalekar 1992), and
3. the dependence of the
slope of the Baldwin effect on the ionization potential of the
emitting ions (Espey \& Andreadis 1999; Green et al. 2001;
Dietrich et al. 2002; Kuraszkiewicz et al. 2002). Non isotropic
continuum emission (Netzer 1987), and the intrinsic Baldwin effect
(Pogge \& Peterson 1992; also review in Osmer \& Shields 1999, \S
5) can produce some of the observed scatter in the Baldwin
effect for the various lines.

Significant correlations also exist among various optical emission
line properties in AGN, including the [O III] and Fe~II strength,
and the \Hb~FWHM, which form part of Eigenvector 1 (EV1) in the
comprehensive study of Boroson \& Green (1992, hereafter BG92).
Boroson, Persson \& Oke (1985) speculated that the underlying
physical parameter which drives these correlation is the relative
accretion rate, \Ledd, a speculation which received strong support
through recent observations (see review in Laor 2000).
Furthermore, Wills et al. (1999) found that various UV emission
properties, including the \C~EW, are correlated with the optical
EV1 parameters, which suggests that the \C~EW may also be largely
driven by \Ledd. The Baldwin effect may then just be a secondary
correlation induced by the tendency of more luminous AGN to have a
higher \Ledd. The possible role of \Ledd\ in the Baldwin effect
was already mentioned by Brotherton \& Francis (1999). Later,
Wilkes et al. (1999) noted that narrow line AGN are outliers to
the Baldwin effect, but the meaning of this result was not
interpreted at that time (see \S 4 here).  Most recently Shang et
al. (2003) suggested an indirect indication, based on spectral
principle component analysis, that \Ledd\ may contribute to the
scatter in the Baldwin effect.

Recent studies established that reasonably accurate estimates of
the black hole mass ({\Mbh}) can be obtained in AGN based on the
continuum luminosity and {\Hb} FWHM (\Mbh$\propto
L^{1/2}$(\Hb~FWHM)$^2$, and thus \Ledd$\propto
L^{1/2}$(\Hb~FWHM)$^{-2}$, e.g. Laor 1998). This opens up the
possibility to test directly whether the Baldwin effect is driven
by {\Ledd}, which is the main point of this paper. The data set
and measurement procedure are described in \S 2, the correlation
analysis is presented in \S 3, and the main results are summarized
in \S 4.

\section{The measurements}

For the purpose of this analysis we use the BG92 sample which
includes the 87 $z<0.5$ AGN from the Bright Quasars Survey (BQS;
Schmidt \& Green 1983). This sample extends in luminosity from
Seyfert galaxies with $\nu L_{\nu}=3.3\times 10^{43}$~erg~s$^{-1}$
(calculated using the continuum fluxes in Neugebauer et al. 1987,
assuming $H_0=80$~km~s$^{-1}$~Mpc$^{-1}$, $\Omega_0=1.0$), to
luminous quasars at $\nu L_{\nu}=1.4\times 10^{46}$~erg~s$^{-1}$,
where $\nu L_{\nu}$ is
calculated at 3000~\AA. This is a complete and well defined
sample, selected based on (blue) color and (point like)
morphology, independently of the emission line strengths. It is
also the most thoroughly explored sample of AGN, with a wealth of
high quality data at most wave bands. The EV1 correlations were
established by BG92 using this sample, and the ability to obtain
reasonably accurate estimates of \Mbh\ (and thus \Ledd) was
demonstrated for this sample (Laor 1998), thus making it an
optimal sample to explore the possible origin of the Baldwin
effect.

Archival UV spectra of the \C\ region are available for 85 of the 87 BG92
objects. The {\it
HST} archives contain UV spectra of 47 of the objects, which were
obtained by the Faint Object Spectrograph (FOS); the UV spectra
for the remaining 38 objects with no {\it HST} spectra
were obtained from the {\it IUE} archives (see Table~1).
An average spectrum, weighted by the S/N ratio,
was calculated when more than one archival spectrum was
available. This averaging should decrease the scatter in the Baldwin
effect induced by variability, as demonstrated in the intrinsic Baldwin effect.

Three of the archival spectra (PG~0934+013, PG~1004+130 and
PG~1448+273) did not have a sufficient S/N to measure the \C~EW,
and in one object (PG~1700+518) \C\ is heavily absorbed (e.g. Laor
\& Brandt 2002), leaving a sample of 81 objects with a measurable
\C~EW. The \C\ region of PG~0043+039 is also rather heavily
absorbed, and its measured \C~EW is probably just a lower limit.
The continuum flux uncertainty of the {\it HST} spectra, which
were obtained without the flux density uncertainty from the {\it
HST} archives, was estimated using the average standard deviation
of the flux at two windows: 1450~\AA\ - 1470~\AA, and 1710~\AA\ -
1730~\AA\ (rest frame)\footnote{Lack of data or strong systematic
features necessitated the following changes to one or both
windows; PG~0003+158: 1465~\AA\ - 1485~\AA, PG~1114+445: 1465~\AA\
- 1485~\AA, and 1730~\AA\ - 1750~\AA; PG~1211+143, PG~1404+226 and
PG~1440+356: only the 1710~\AA-1730~\AA\ window used; PG~1415+451:
used only 1700~\AA\ - 1720~\AA; PG~1543+489: 1600~\AA\ - 1620~\AA;
PG~1612+261: 1470~\AA\ - 1490~\AA; PG~2304+042: 1680~\AA\ -
1700~\AA.}. The rest frame wavelengths were calculated using
redshifts determined from the peak of the [O~III]~$\lambda 5007$
line (kindly provided by T. Boroson, private communication). The
wavelength dependence of the flux density uncertainty (of the {\it
HST} spectra) was assumed to scale with the square root of the
ratio of the flux density to continuum flux density. The {\it IUE}
spectra were obtained with the flux density uncertainty from the
{\it IUE} archives.

The spectra were corrected for Galactic reddening using the {\it
E(B$-$V)} values from  Schlegel, Finkbeiner \& Davis (1998, as
listed in the NASA/IPAC Extragalactic Database), and the reddening
law of \citet{Seaton79}. A local power-law continuum was fit to
each spectrum between $\sim 1470$~\AA\ and  $\sim 1620$~\AA, and
the \C\ line emission was fit as a sum of three Gaussians, using
the procedure described in Laor et al. (1994, \S 3 and the
appendix there). Wavelength regions suspected to be affected by
intrinsic or Galactic absorption were excluded from the fit
(Laor \& Brandt 2002). The purpose of the line fit is not to
decompose the line to possible components, but rather to obtain a
smooth realization of the line profile, which is likely to yield
more accurate values for the line width and EW.

Table 1 presents our measured rest-frame EW of the best-fitting
models to the \C\ profile, together with the estimated errors.
The errors were estimated
by repeating the model fits with the power-law continuum displaced
upward and downward by $1\sigma$. As expected, the
typical errors associated with {\it IUE} spectra are significantly
higher than those of the {\it HST} spectra. One should note that
larger systematic errors could be associated with the placement of the
continuum windows. In particular, the presence of very broad weak
wings is difficult to detect, although they may have a non-negligible EW.
Table 1 also lists $\nu L_{\nu}$ calculated at 3000~\AA\ for each object,
and the estimated \Ledd.


\begin{table*}
\centering
\begin{minipage}{175mm}
\caption{The \C\ EW, $\nu L_{\nu}$, and \Ledd\ of the 81 BQS
quasars.} \label{fit} \setlength{\tabcolsep}{0.75ex}
\begin{tabular}{@{}lr@{$\pm$}rrrp{2ex}lr@{$\pm$}rrrp{2ex}lr@{$\pm$}rrr@{}}
\hline Object  & \multicolumn{2}{c}{EW\footnote{In units of \AA.
EW values with a decimal point are based on {\it HST} spectra,
while the integer rounded values are based on {\it IUE}
spectra.\label{fn:1a}}} & \multicolumn{1}{c}{$\nu
L_{\nu}$\footnote{In units of $\log (\nu L_{\nu}/10^{44}$), where
$\nu L_{\nu}$ is measured in erg~s$^{-1}$ at rest frame
3000\AA.\label{fn:1b}}}&\multicolumn{1}{c}{\Ledd\footnote{ Log
\Ledd.\label{fn:1c}}} & & Object &
\multicolumn{2}{c}{EW\fnrepeat{fn:1a}} & \multicolumn{1}{c}{$\nu
L_{\nu}$\fnrepeat{fn:1b}}&\multicolumn{1}{c}{\Ledd\fnrepeat{fn:1c}}
& & Object & \multicolumn{2}{c}{EW\fnrepeat{fn:1a}} &
\multicolumn{1}{c}{$\nu L_{\nu}$\fnrepeat{fn:1b}}&\multicolumn{1}{c}{\Ledd\fnrepeat{fn:1c}} \\
\hline 0003+158    &   63.5    &   4.6 &   1.877   &  $-$0.358  &&
1115+407    &   25.9    &   4.2 &   0.547   &  $-$0.139  &&
1416$-$121    &   168.1   &   40.2    &   1.337
&  $-$0.845  \\
0003+199    &   60.1    &   2.6
&   0.059   &  $-$0.342  &&  1116+215    &   40.5    &   2.9
&   1.468   &  $-$0.139  &&  1425+267    &   64.8    &   10  &   1.219   &
$-$1.280   \\
0007+106    &   59  &   5
&   0.770    &  $-$0.972  &&  1119+120    &   29    &   5
&  $-$0.001  &  $-$0.462  &&  1426+015    &   32    &   2
&   0.985   &  $-$1.117  \\
0026+129    &   19.3    &   3.9
&   1.067   &   0.053   &&  1121+422    &   41.7    &   4.1
&   0.806   &  $-$0.232  &&  1427+480    &   53.2    &   3.7
&   0.815   &  $-$0.344  \\
0043+039    &   5.4 &   3.7 &   1.485   &  $-$0.648  &&  1126$-$041
&   30    &   7 &   0.344   &  $-$0.434  &&  1435$-$067    &   39 &
7
&   1.069   &  $-$0.412  \\
0049+171    &   203   &   73    &  $-$0.109  &  $-$1.437  &&
1149$-$110    &
   82    &   20    &  $-$0.006  &  $-$0.916  &&  1440+356    &   30.1    &
  1.4 &   0.503   &  $-$0.013  \\
0050+124    &   29.9    &   1.5
&   0.582   &   0.162   &&  1151+117    &   26.6    &   7.1
&   0.815   &  $-$0.801  &&  1444+407    &   17.9    &   1.1
&   1.217   &  $-$0.122  \\
0052+251    &   119.0
&   10.5    &   1.104   &  $-$0.822  &&  1202+281    &   290.0
&   31.3    &   0.590    &  $-$1.053  &&  1501+106    &   64    &   1
&   0.491   &  $-$1.172  \\
0157+001    &   43    &   8
&   0.926   &  $-$0.261  &&  1211+143    &   55.7    &   1.8
&   1.063   &   0.051   &&  1512+370    &   84.3    &   7.2
&   1.483   &  $-$0.867  \\
0804+761    &   45    &   3
&   1.233   &  $-$0.300    &&  1216+069    &   64.5    &   4.4
&   1.527   &  $-$0.609  &&  1519+226    &   68    &   16    &   0.647   &
  $-$0.311  \\
0838+770    &   50    &   10    &   0.679   &  $-$0.493  &&  1226+023    &
   23.0  &   0.7 &   2.045   &  $-$0.012  &&  1534+580    &   79    &   6
&  $-$0.337  &  $-$1.565  \\
0844+349    &   28    &   5
&   0.461   &  $-$0.479  &&  1229+204    &   48    &   3
&   0.381   &  $-$0.804  &&  1535+547    &   27.6    &   1.7
&  $-$0.182  &  $-$0.373  \\
0921+525    &   186   &   11    &  $-$0.415  &  $-$0.802  &&  1244+026    &
   17    &   4
&   0.031   &   0.235   &&  1543+489    &   25.6    &   1.4
&   1.394   &   0.369   \\
0923+129    &   93    &   13  &  $-$0.250   &  $-$0.665  &&  1259+593    &
15.3    &   2.5
&   1.834   &  $-$0.085  &&  1545+210    &   90.5    &   10.5    &   1.421
&  $-$0.925  \\
0923+201    &   28    &   6 &   1.141   &  $-$1.134  &&  1302$-$102
&   13.1    &   1.6 &   1.850    &  $-$0.080   &&  1552+085    & 47
&   16  &   0.585   &
0.040    \\
0947+396    &   55    &   4   &   0.802   &  $-$0.909  &&  1307+085    &
71.2    &   8.5
&   1.071   &  $-$0.651  &&  1612+261    &   94.6    &   13.9    &   0.699
&  $-$0.395  \\
0953+414    &   54.9    &   5   &   1.473   &  $-$0.196  &&  1309+355    &
33.5    &   5.5 &   0.915   &  $-$0.421  &&  1613+658    &   54    &   3
&   0.676   &  $-$1.457  \\
1001+054    &   34.9    &   4.6 &   0.806   &  $-$0.020   &&
1310$-$108    &   78    &   16    &  $-$0.243
&  $-$1.183  &&  1617+175    &   34    &   7   &   1.030    &  $-$0.880   \\
1011$-$040    &   25  &   5 &   0.225   &  $-$0.146  &&  1322+659 &
52.6    &   3.4 &   0.847   &  $-$0.409  &&  1626+554    & 45.6
&   7.6
&   0.612   &  $-$0.940   \\
1012+008    &   23    &   6   &   0.931   &  $-$0.320   &&  1341+258    &
62    &   20  &   0.304   &  $-$0.756  &&  1704+608    &   34.8    &   5.2
&   1.606   &  $-$0.772  \\
1022+519    &   38  &   11    &  $-$0.479  &  $-$0.600    &&  1351+236    &
   101   &   48    &  $-$0.351  &  $-$1.748  &&  2112+059    &   25.5    &
  3.5 &   2.131   &   0.116   \\
1048$-$090    &   91    &   50    &   1.524   &  $-$0.679  &&
1351+640    &
   43.3    &   4.4
&   0.780    &  $-$1.058  &&  2130+099    &   47    &   3
&   0.619   &  $-$0.367  \\
1048+342    &   46    &   17    &   0.735   &  $-$0.687  &&  1352+183    &
   45.1    &   6.5
&   0.851   &  $-$0.629  &&  2209+184    &   54    &   21    &   0.428   &
  $-$1.353  \\
1049$-$006    &   67.0  &   8.8 &   1.540    &  $-$0.630   &&
1402+261    &   30.3    &   2.8 &   1.044   &   0.018   &&
2214+139    &   45    &   4
&   0.462   &  $-$1.027  \\
1100+772    &   84.0  &   4.9
&   1.544   &  $-$0.749  &&  1404+226    &   23.3    &   3.4
&   0.126   &   0.232   &&  2251+113    &   66.0  &   3.5
&   1.634   &  $-$0.363  \\
1103$-$006    &   37.2    &   9   &   1.575   &  $-$0.737  &&
1411+442    & 56.9    &   18  &   0.520    &  $-$0.535  && 2304+042
&   176
&   48    &  $-$0.133  &  $-$2.018  \\
1114+445    &   55.0  &   4.1
&   0.669   &  $-$0.927  &&  1415+451    &   57.3    &   3.9
&   0.399   &  $-$0.579  &&  2308+098    &   81.5    &   6.8
&   1.616   &  $-$0.936  \\
\hline
\end{tabular}
\end{minipage}
\end{table*}

\section{The correlation analysis}

Figure 1 (top panel) presents the Baldwin effect for the 81 BQS
quasars measured above. The \C~EW are taken from Table 1, and for
the luminosity we use  $\nu L_{\nu}$ at 3000~\AA\ (see \S 2). To
verify that our sample is not biased in any way, we also present
in Fig.~1 data from two other studies, 454 Large Bright Quasar
Survey objects\footnote{We have ignored two upper limits in the
tabulation of \citet{For01}, and several obvious typographical
flux errors, in their total list of 488 objects.} analyzed by
\citet{For01}, and 125 pre-COSTAR AGN
observations analyzed by \citet{Kur02} (which overlap some of our
objects). These data are available in public electronic form. The
luminosities of these objects were derived from the specified flux
densities using the same cosmology assumed for the BQS quasars.
The trend and scatter of the
Baldwin effect for the BQS quasars appears to be very similar to
those displayed by the \citet{For01} and \citet{Kur02} samples.
The Spearman rank-order correlation coefficient for the BQS
Baldwin effect is $r_s=-0.154$. The flattening of the
Baldwin relation at low luminosity is clearly seen in Fig.~1 (see
discussion in \S 4.2 of Osmer \& Shields 1999). The non-linearity
of the Baldwin effect (on log-log scale) suggests that the
Spearman rank-order correlation coefficient, which tests for a
monotonic relation, is a more suitable statistical test here,
compared to the Pearson correlation coefficient, which tests the
significance of a linear relation.

Table 2 presents the results of a correlation analysis of the
\C~EW reported in Table 1 with all the optical emission line
parameters from BG92 (table 2 there), supplemented by
$\alpha_{ox}$ from
Brandt, Laor \& Wills (2000), and \Ledd\ and \Mbh\ as estimated
from the optical luminosity and the \Hb~FWHM (equation 3 in Laor
1998). For the sake of brevity we present only the most
significant correlations (those where Pr~$<5\times 10^{-4}$). To
explore the effect of the generally lower S/N {\it IUE} spectra on
the strength of the correlations, we also present in Table 2 the
correlations obtained based only on the 46 objects with the
generally higher S/N {\it HST} spectra. The correlations for the
{\it HST} sample are similar, which
suggests that the \C~EW measurement uncertainty is not a major source
error.

The strongest correlations of the \C~EW are with $\alpha_{ox}$, and
with parameters
related to the strength of the \Ox\ and \Fe\ lines, and with the
\Hb~FWHM, which are part of the BG92 EV1 correlations.
These correlations confirm earlier
results on the inverse relation of the \C~EW and the \Fe\ strength (Marziani
et al. 1996; Wang et al. 1996), the \C~EW EV1 correlations found
by Wills et al. (1999, table 1 there) for a smaller sample of 22
BQS quasars (part of our sample), and the earlier
results on the \C~EW correlation with $\alpha_{ox}$ (see \S 1).
The correlation with $\alpha_{ox}$ is consistent with the suggestions
that the Baldwin effect is driven by a softening of the ionizing
continuum (which decreases $\alpha_{ox}$), with increasing
luminosity (Korista et al. 1998). The inverse correlation of the
\C~EW  and the \Fe\ strength may be due to optical
depth effects. A large optical depth in the \Fe\ emitting region
strengthens the optical \Fe\ emission by converting UV~\Fe\ emission
to optical \Fe\ emission (Netzer \& Wills 1983, Shang et al. 2003).
In contrast, a large optical depth in the \C\
emitting region results in collisional suppresses of this line
(Ferland et al. 1992). Thus, the observed inverse correlation could
be produced if the optical depths of the \Fe\ and \C\ emitting regions
are related.

Objects with  $\alpha_{ox}<-2$, or
`Soft X-Ray Weak' AGN, generally show intrinsic broad \C\
absorption (Brandt et al. 2000), whose strength is correlated with
luminosity (Laor \& Brandt 2002). Incomplete correction for this
absorption may induce the observed trend of decreasing \C~EW with
decreasing $\alpha_{ox}$. To avoid absorption biases, we repeated
the \C~EW vs. $\alpha_{ox}$ correlation excluding the 12 objects
with $\alpha_{ox}<-1.75$ where there is potentially significant absorption.
This gave essentially the same result
($r_S=0.498$, vs. $r_S=0.525$ for the complete sample), indicating
there is no significant absorption bias.

The interesting new result in Table 2 (Fig.~1, middle panel) is
the strong correlation of the \C~EW with the \Ledd\ indicator
(i.e. $L^{1/2}$(\Hb~FWHM)$^{-2}$), where $r_S=-0.581$ (null
probability of Pr = $1.3\times 10^{-8}$). This is not a simple
consequence of the separate correlations of the \C~EW with $L$ and
with the \Hb~FWHM, as for example, the correlation with \Mbh\ (i.e.
$L^{1/2}$(\Hb~FWHM)$^2$) is insignificant ($r_S=0.21$, Pr =
0.05). However, does the combination of $L$ and Hb~FWHM in the
form $L^{1/2}$(\Hb~FWHM)$^{-2}$ produces the tightest correlation?
To answer that we searched for the largest $r_S$ in a correlation of
log~\C~EW with the linear combination log~$L+a\log$(\Hb~FWHM), where
$a$ is a free parameter. We found a maximum correlation of
$r_S=-0.588$ at $a=-5.3$, close to the combination of $L$ and
\Hb~FWHM which provides \Ledd\ ($a=-4$). Thus, \Ledd\ may be
the primary parameter which drives the Baldwin effect for \C. This implies
that the large scatter in the Baldwin effect is produced by the range of
\Ledd\ at a given $L$, as suggested by Shang et al. (2003). However, our
results disagree with the suggestion of Shang et al. that the \C~EW is
primarily dependent on $L$, rather than \Ledd.

Is \Ledd\ the {\em only} parameter which controls the \C~EW?
Since the EV1 correlations also appear to be driven by \Ledd\
(BG92; Boroson 2002), their correlation with the \C~EW could be
interpreted as secondary correlations. This can be tested with a partial
correlation analysis, which yields an average $r_S=0.42$ for the
correlation of \C~EW with $\alpha_{ox}$ and the EV1 parameters,
keeping \Ledd\ fixed, and also for the correlation of \C~EW with
\Ledd, keeping EV1 and $\alpha_{ox}$ fixed. Although these
correlations are weaker than the original ones (Table 2), they are
still significant (Pr $\sim 10^{-4}$).  The significance of the
partial correlations suggests there is a true scatter in the BLR
properties at a fixed \Ledd, though it may also be induced by the
inaccuracy of our \Ledd\ indicator.

The significance of the partial correlations implies that the
scatter in the \C~EW vs. \Ledd\ relation can be reduced by
including a third parameter (in addition to $L$ and the \Hb~FWHM,
which form \Ledd). The parameter which most significantly improves
this correlation is the \Ox~$\lambda 5007$ EW (Fig. 1, lower
panel), which yields $r_s=-0.722$ (using $\alpha_{\rm ox}$
instead of \Ox~$\lambda 5007$ EW as a
third parameter yields $r_s=-0.667$, and using the \Fe/\Hb\ flux ratio
yields $r_s=-0.674$). Similarly, the most significant fourth
parameter is $\alpha_{\rm ox}$ (i.e. correlating log \C~EW  vs. a
linear combination of log \Ledd, log \Ox~$\lambda 5007$ EW, and
$\alpha_{\rm ox}$), which yields a small improvement to
$r_s=-0.753$. Finally, the inclusion of  the \Fe/\Hb\ flux ratio as
a fifth parameter yields a slight improvement to $r_s=-0.767$.

Some of the scatter in the \C~EW correlations may be induced by a
spread in the BLR covering factors among AGN. A plausible way to
overcome such a spread is to use the \C~EW/\Hb~EW ratio in the
above correlations, instead of the \C~EW.  This yields significantly
{\em lower} correlations
(e.g. the correlations with \Ledd\ and $\alpha_{\rm ox}$ go down to
$-$0.391 and 0.391 from $-$0.581 and 0.525, respectively). This
may suggest that the \Hb~EW is not a good indicator of the BLR
covering factor, but is rather modulated by other BLR properties
which do not modulate the \C~EW.

\begin{table}
\begin{minipage}{84mm}
\caption{The main \C~EW correlations.} \label{tcorr}
\begin{tabular}{@{}lrc@{}}
\hline Variable Name\footnote{The prefix `R' indicates the ratio
of the variable to that of \Hb\ (BG92), e.g. R~\Fe~EW corresponds
to \Fe~EW/\Hb~EW.} & $r_s$\footnote{The top
value for each variable is obtained for the complete sample of 81
objects, the bottom value corresponds to the sub-sample of 46
objects with {\it HST} spectra.\label{fn:2b}}
&  Pr\fnrepeat{fn:1b}\\
\hline
$\nu L_{\nu}(3000$\AA)\dotfill &$-$0.154 & 1.71$\times10^{-01}$ \\
&$-$0.018 & 9.08$\times10^{-01}$ \\
\Ledd\dotfill  &  $-$0.581    &   1.31$\times10^{-08}$    \\
&$-$0.642 & 1.53$\times10^{-06}$ \\
$\alpha_{\rm ox}$\dotfill   &   0.525    &   4.87$\times10^{-07}$    \\
& 0.463 & 1.18$\times10^{-03}$ \\
\Ox~$\lambda 5007$~EW\dotfill   &   0.624 &   4.71$\times10^{-10}$    \\
& 0.708 & 3.67$\times10^{-08}$ \\
\Fe~EW\dotfill   &  $-$0.518    &   7.49$\times10^{-07}$    \\
&$-$0.536 & 1.24$\times10^{-04}$ \\
\Hb~FWHM\dotfill   &   0.427    &   7.03$\times10^{-05}$    \\
& 0.510 & 2.92$\times10^{-04}$ \\
R~\Ox~$\lambda 5007$ peak height\dotfill  &   0.624
&   4.78$\times10^{-10}$    \\
& 0.647 & 1.20$\times10^{-06}$ \\
R~\Fe~EW\dotfill  &  $-$0.626   &   4.02$\times10^{-10}$    \\
&$-$0.698 & 6.94$\times10^{-08}$ \\
R~\Ox~$\lambda 5007$~EW\dotfill    &   0.471 &   9.23$\times10^{-06}$    \\
& 0.494 & 4.89$\times10^{-04}$\\
\hline
\end{tabular}
\end{minipage}
\end{table}

\begin{figure}
\includegraphics[width=85mm]{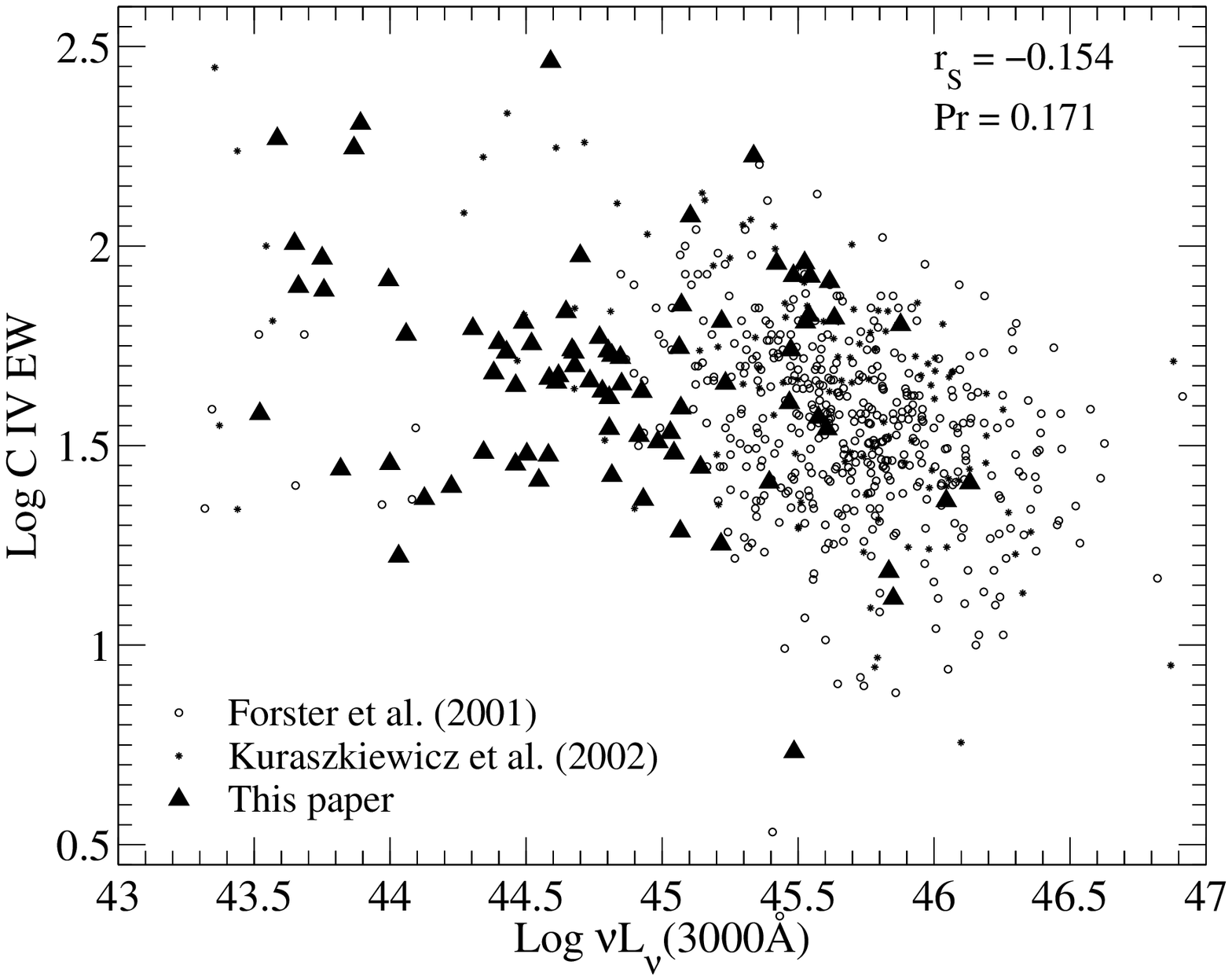}
\includegraphics[width=85mm]{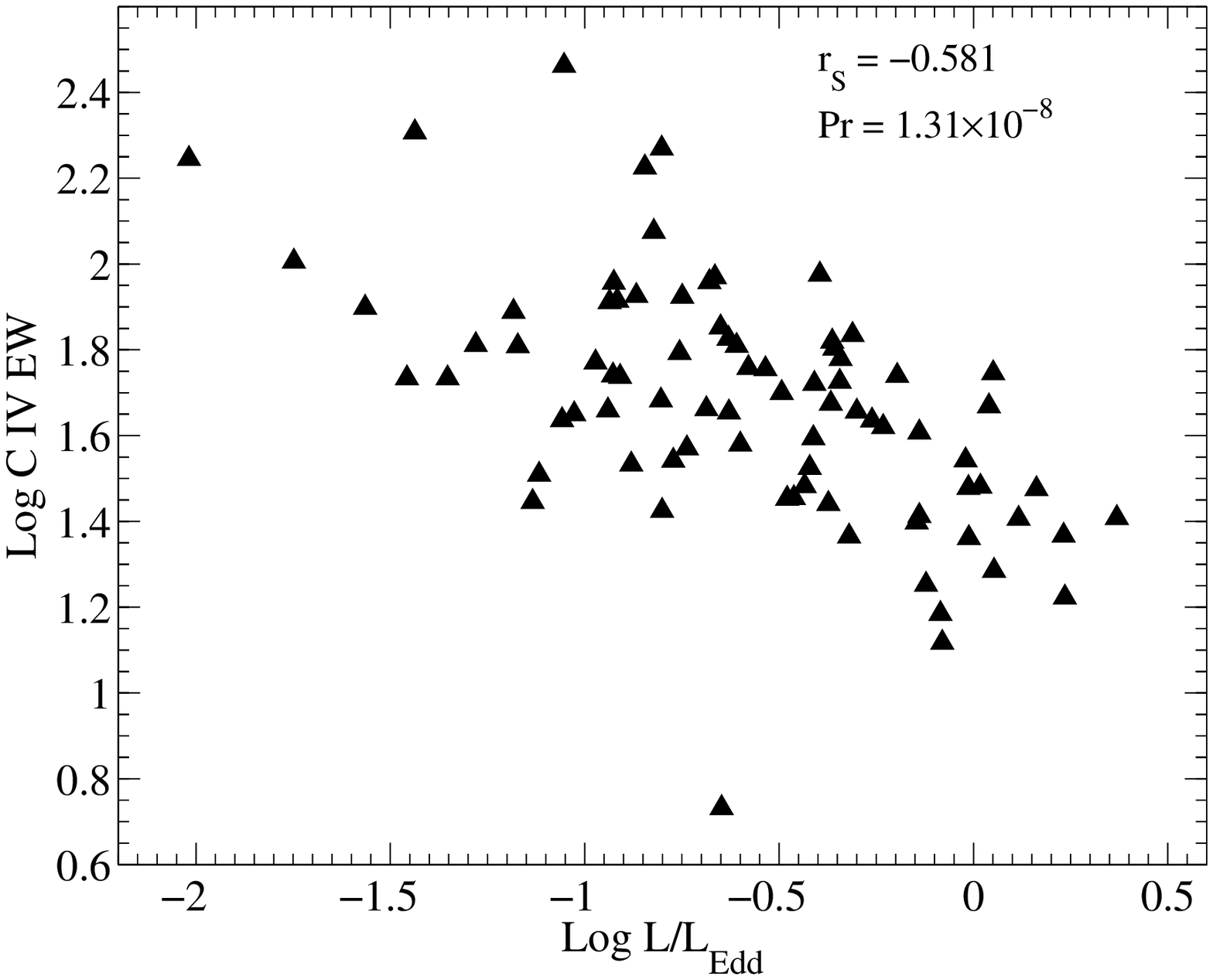}
\includegraphics[width=85mm]{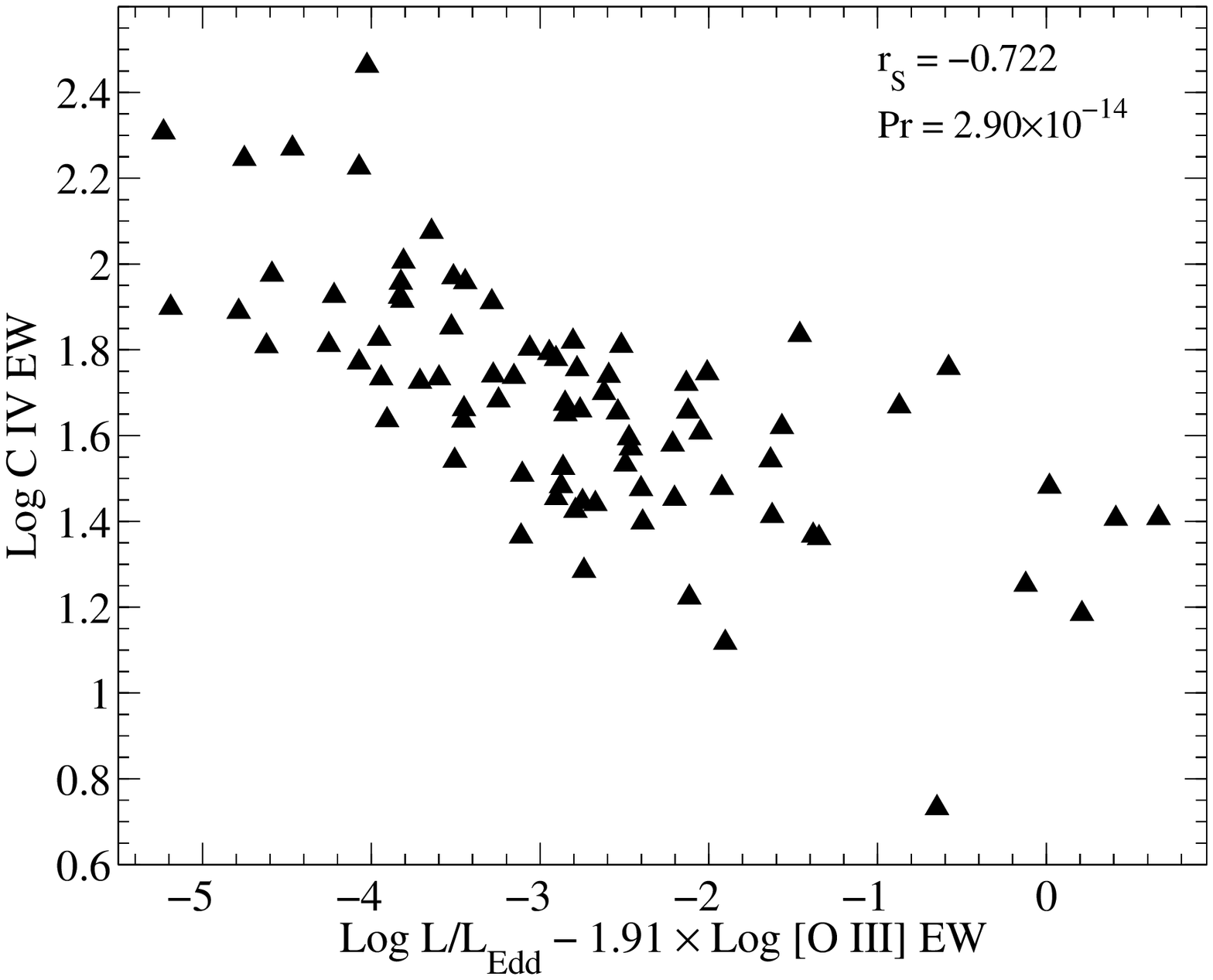}
\caption{The main \C~EW correlations discussed in this paper. The
Spearman rank-order correlation coefficient ($r_S$) and the null
probability (Pr) are indicated at each panel. Top panel: The
Baldwin effect for the 81 BQS quasars, together with the
\citet{For01} and \citet{Kur02} samples. All samples are similarly
distributed. Middle panel: The correlation of the \C~EW with
{\Ledd}. Note the significant increase in $r_S$. Lower panel: The
correlation with the addition of \Ox~EW as a third parameter,
which further increases $r_S$. } \label{fcorr}
\end{figure}

\section{Conclusions}

The purpose of this paper is to obtain some indications for the physical parameters
which drive the Baldwin effect in \C\ through correlation analysis.
We use archival {\it HST} and {\it IUE} \C\ spectra of sufficient quality
available for 81 of the 87 BQS quasars, together with the optical
emission line parameters from BG92, and $\alpha_{\rm ox}$ from Brandt
et al. (2000).

We find that a major source of scatter in the Baldwin effect is the
\Hb~FWHM. Its inclusion as a second parameter to $L$, in the form
$L^{1/2}\times$(\Hb~FWHM)$^{-2}$, which is proportional to \Ledd,
leads to a significant increase in the correlation strength
(from $r_s=-0.154$ to $r_s=-0.581$). This indicates that the Baldwin
effect is a secondary relation, which is induced by the stronger relation
of the \C~EW and \Ledd, and the tendency of high $L$ AGN to have a
higher \Ledd. This also explains why NLS1s, which are high \Ledd\
AGN at a low $L$, have much lower \C~EW than expected for their $L$
(as found by Wilkes et al. 1999).

We do not know the physical mechanism responsible for the reduction
in the \C~EW with increasing \Ledd. However, we find that
the scatter in the Baldwin effect can be further reduced by including
either the \Ox~EW, the relative \Fe\ strength, or $\alpha_{\rm ox}$,
as a third parameter. The \Fe\ strength may be an indicator of the
BLR optical depth, and $\alpha_{\rm ox}$ may be an indicator of the ionizing
continuum shape, which are known theoretically to affect the \C~EW.
The physical mechanism implied by the correlation
with the \Ox~EW remains to be understood.
At least some of the remaining scatter can be
produced by a non-isotropic, or a time variable continuum source.

The results presented here can be tested with larger samples of AGN.
In particular, it will be interesting to explore the extension to high
$z$ AGN, where the \Hb\ region is observable in the IR (e.g. McIntosh et al. 1999;
Yuan \& Wills, 2003; Shemmer et al. 2004).
In addition, different lines show
different slopes for the Baldwin effect, and it will be interesting
to explore what are the primary physical parameters which drive other
emission lines.
\section*{acknowledgments}
We thank the referee for some very helpful remarks. This research has made use of
the NASA/IPAC Extragalactic Database
(NED), which is operated by the Jet Propulsion Laboratory,
California Institute of Technology, under contract with the
National Aeronautics and Space Administration.

\end{document}